# High-Q surface electromagnetic wave resonance excitation in magneto-photonic crystals for super-sensitive detection of weak light absorption in near-IR

O.V. BOROVKOVA,[1,*] D.O. IGNATYEVA,[1,2] S.K. SEKATSKII,[3] A. KARABCHEVSKY,[4,**] V.I. BELOTELOV[1,2]

[1]*Russian Quantum Center, Skolkovo, Moscow Region 143025, Russia*
[2]*Lomonosov Moscow State University, Moscow 119991, Russia*
[3]*Laboratoire de Physique de la Matière Vivante, IPHYS, EPFL, 1015 Lausanne, Switzerland*
[4]*School of Electrical and Computer Engineering, Ben-Gurion University of the Negev, Beer-Sheva 8410501, Israel*
*Corresponding authors: \*o.borovkova@rqc.ru, \*\*alinak@bgu.ac.il*

**Mid-infrared spectrum can be recorded from almost any material making the mid-infrared spectroscopy an extremely important and widely used sample characterization and analytical technique. However, sensitivity photoconductive detectors operate primarily in the near-infrared (NIR) but not in the mid-infrared making the NIR more favorable for accurate spectral analysis. Although absorption cross-section of vibrational modes in near-infrared is orders of magnitude smaller compared to the fundamental vibrations in mid-infrared, different concepts were proposed to increase the detectability of weak molecular transitions overtones. Yet, the contribution of magneto-photonic structures in near-infrared absorption effect has never been explored so far. Here we propose high-Q magneto-photonic structures for a super-sensitive detection of weak absorption resonances in near-infrared. We analyze the contributions of both magnetic and non-magnetic photonic crystal (PC) configurations to the detection of weak molecular transitions overtones. Our results constitute an important step towards development of highly sensitive spectroscopic tools based on high-Q magneto-photonic sensors. © 2019 Optical Society of America**

http://dx.doi.org/10.1364/AO.99.099999

## 1. INTRODUCTION

The absorption spectroscopy in the mid-infrared (mid-IR) range is widely used to investigate and detect the organic molecules [1]. This technique allows detecting molecules in mixtures of substances without prior separation but rather due to the well-defined spectral lines associated with fundamental vibrations. Spectral analysis is particularly important for fixing minor changes in the composition of proteins by studying the fingerprints (identification) of the molecular bonds, for instance, the peptide bonds in proteins [2, 3]. However, highly sensitive photoconductive detectors function in the near-infrared (NIR) but not in the mid-IR. Also affordable materials, such as glass and quartz, are transparent for NIR radiation and thus can be used as cell windows, focusing lenses, and optical fibers.

The circumstance that the absorption cross sections of the higher harmonics of vibrational transitions are several orders of magnitude smaller than those of the fundamental transitions, and have a low signal-to-noise ratio in the absorption spectrum, makes it difficult to use absorption spectroscopy methods in the near-IR range for the identification of molecules. Recently, A. Karabchevsky and her group proposed to explore the high-order vibrational transitions (overtones), lying in the NIR range, on glass waveguides [4]. Earlier, giant absorption of light was observed and explained as a switch from ballistic to diffusive regime in borosilicate waveguides [4] and silicate microfibers [5]. Several concepts were proposed to explore the interaction of electromagnetic radiation with overtone transitions which are forbidden transitions in the dipole approximation, such as e.g. glass waveguide described above, silicon rib waveguides [6], planar and cylindrical waveguides [7]. One of the most promising methods to increase the sensitivity of the infrared absorption spectroscopy is the two orders of magnitude amplification of the electric field when the plasmon resonances are excited [8-12].

High sensitivity of the surface plasmon resonance to the refractive index of the external medium underlies widely used surface plasmon resonance (SPR)-based sensors [13, 14] which detect the smallest variations of the refractive index of analyzed substances as the shift of the optical resonance associated with SPP excitation. However, SPP on a smooth metal surface provides rather low-Q resonance so there are many ways of the system improvement. For example, optical resonances in nanostructured plasmonic materials [10-12] are shown to be perspective for sensing applications. However, utilization of such complex structures is challenging so another way is to design the layered structure with high-Q resonance. It has been demonstrated earlier the plasmonic structures exhibit a very high sensitivity to the small nonlinear and gyrotropic components of the permittivity tensor [15-17]. This gives rise to several orders enhancement of the magneto-optical effects in magnetoplasmonic structures [18-21]. Magnetooptical response, i.e. light polarization rotation or intensity modulation, enhanced by several orders of magnitude near the surface resonance is more sensitive than the reflectance spectra measured in non-magnetic SPR sensors. Therefore, a significant improvement of sensitivity and the resolution of magneto-optical surface resonance-based sensors, as well as the signal-to-noise ratio was achieved by utilization of the magnetic layers in the plasmonic structure [22-27+список]. However, a physical limit of the SPR sensing is associated with the low quality factor arising from light absorption by metals. Ferromagnetic metals or dielectrics possess even higher losses than non-magnetic ones. In traditional sensors, where the variations of the real part of the refractive index are measured, utilization of all-dielectric structures supporting high-quality quasisurface wave resonances instead of plasmonic structures was demonstrated to increase the signal-to-noise ratio and sensitivity by an order of magnitude [28, 29]. Even higher sensitivity was demonstrated for magnetophotonic structures where the ultra-high quality-factor magneto-optical resonances were excited [29-33].

In this paper, for the first time, we propose the utilization of all-dielectric magnetophotonic structures for sensing of the imaginary part of the permittivity having very weak resonances in NIR. We show that excitation of high-Q resonances of quasi-surface waves in all-dielectric structures is promising for measurements of the vibrational transitions of organic molecules for several reasons. First, the small absorption peaks of the studied substances in NIR is the only source of absorption in all-dielectric transparent structures. So, high Q of the resonances is responsible for a high sensitivity to the losses produced by absorption peaks. Then, magnetooptical Kerr effect is known [34] to depend on the structure absorption significantly. Thus, even the minute peaks in the analyzed molecules absorption spectra that refer to the higher harmonics of the vibrational transitions could be effectively detected. We would like to point out that the influence of magnetic field on weak absorption by molecular transitions overtones was not studied before.

Here, we propose two Photonic Crystal (PC)-based nanostructures for very high-Q sensing of N-methylacetamide (NMA) molecules. Both structures provide narrow resonances in reflection spectrum corresponding to the absorption peaks typical for NMA molecules [4, 7]. This allows to detect the presence of NMA molecules in analyte with high sensitivity. The first PC nanostructure is non-magnetic and includes just four pairs of [silicon]/[silicon nitride] and the additional silicon layer. The nanostructure demonstrates an ultranarrow resonance dip in reflectance associated with small absorption peak while illuminated by s-polarized light. It is worth noting, that this structure is simple in terms of fabrication due to the small number of layers. The second structure provides more sensitive detection due to the higher quality of the magneto-photonic resonance in comparison with the non-magnetic PC nanostructure.

## 2. ULTRALONG-PROPAGATION QUASI-SURFACE MODES FOR ABSORPTION PEAK DETECTION

The sensing concept is schematically depicted in Fig. 1. The multilayered stack of PC nanostructure is placed on a prism and illuminated by the focused polychromatic incident light beam. We design the PC structure to support the excitation of the long-range propagating quasi-surface electromagnetic modes. The excited mode represents a guiding mode localized in the terminating layer of the PC structure due to the PC bandgap on the one side and total internal reflection on the other. The parameters of the PC are tuned so that under the near-cutoff conditions it has strongly asymmetric profile with large penetration depth inside of the analyzed substance and therefore exhibits high sensitivity to analyte permittivity. Although the excited mode is guided by its nature, in the sense of field distribution and sensitivity it is very similar to surface plasmon polariton modes, so further this mode is referred to as 'quasi-surface'. Such types of modes were shown [28-30] to combine the advantages of the guided modes - such as high Q, and surface modes - such as large penetration depth and high sensitivity to the optical properties of the external medium.

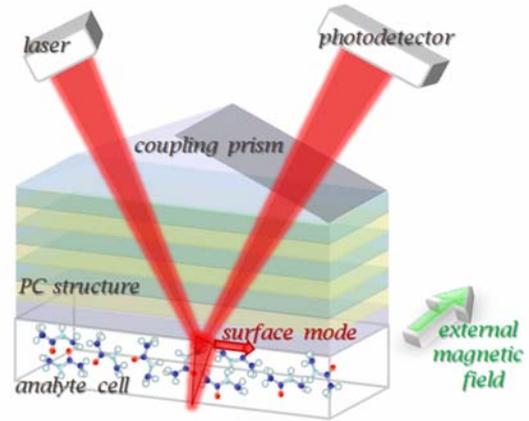

Fig. 1. Schematic representation of the proposed sensor. Incident light illuminates the prism which in turn provides the momentum to excite the quasisurface modes. The surface modes excite the studied organic molecular vibrations overtones in PC-based multilayer sensor which is observed in the reflectance spectra of s-polarized light for non-magnetic structure and in transverse magneto-optical Kerr effect spectra of p-polarized light for magnetic structure.

The spectral position of this mode resonance is tuned to overlap with spectral resonances of the overtone transitions of the studied analyte molecules. We have selected N-methylacetamide (NMA) as analyzed substance since it exhibits a very small absorption peak with magnitude of the extinction coefficient $n''=10^{-4}$ at the wavelength 1.495 um [4,7]. We notice that the excited surface mode in all-dielectric lossless transparent structure is very sensitive to the small absorption peaks arising from the vibration overtones of molecules, since it is the only source of absorption in the structure (in contrast to the plasmonic structures). The presence of the absorption at the peculiar frequencies results in the appearance of the corresponding narrow dips in the reflectance spectra of the PC-based sensor totally vanishing if the absorption is set to zero. Thus, the PC-based structure provides an efficient method of the small absorption peaks measurement.

Further improvement of the discussed structure could be performed if the terminating (guiding) layer of the PC-based structure is magnetic. In this case, transversal magneto-optical Kerr effect $\delta_R$ (TMOKE) could be observed in reflection as the relative change of the reflection coefficient of the PC nanostructure $R$ when the structure is remagnetized in opposite directions *+M* and *-M* via external magnetic field [28]:

$$\delta_R = \Delta R/R = 2(R(\boldsymbol{M})-R(\boldsymbol{-M}))/(R(\boldsymbol{M})+R(\boldsymbol{-M})). \quad (1)$$

Here we denote the absolute change of the reflection coefficient R when the structure is remagnetized as $\Delta R = R(\boldsymbol{M})-R(\boldsymbol{-M})$. TMOKE is significantly enhanced near the excited resonance dip in reflectance,

## 3. ULTRAHIGH-Q RESONANCE IN NON-MAGNETIC PC-BASED STRUCTURE

In order to excite quasi-surface electromagnetic mode at PC-analyte interface, we designed the PC-based structure which consists of 4 pairs of layers of silicon nitride $Si_3N_4$ with the thickness of 222 nm and silicon Si with the thickness of 135 nm, while the thickness of the bottom Si layer is 260 nm. The thickness of the bottom Si layer h determines both an angular position and the quality of the resonance in the reflection spectra, so that $h$=260 nm provides the balance between the quality of the resonance and the typical tolerance of fabrication process about several nm (see Appendix A). The PC structure and electromagnetic field distribution is shown in Fig.2a. The proposed design of the PC non-magnetic sensor provides TE-polarized surface mode at 1.495 um with the evanescent field penetrating into the analyte (NMA). S-polarized light illuminates the facet of the Si equilateral prism with angle of 27° to the normal to the PC surface. Although the PC has a center of the photonic bandgap (BG) at 27° (Fig.2b), due to the finite thickness of the PC the light partially goes through the PC till the bottom silicon layer.

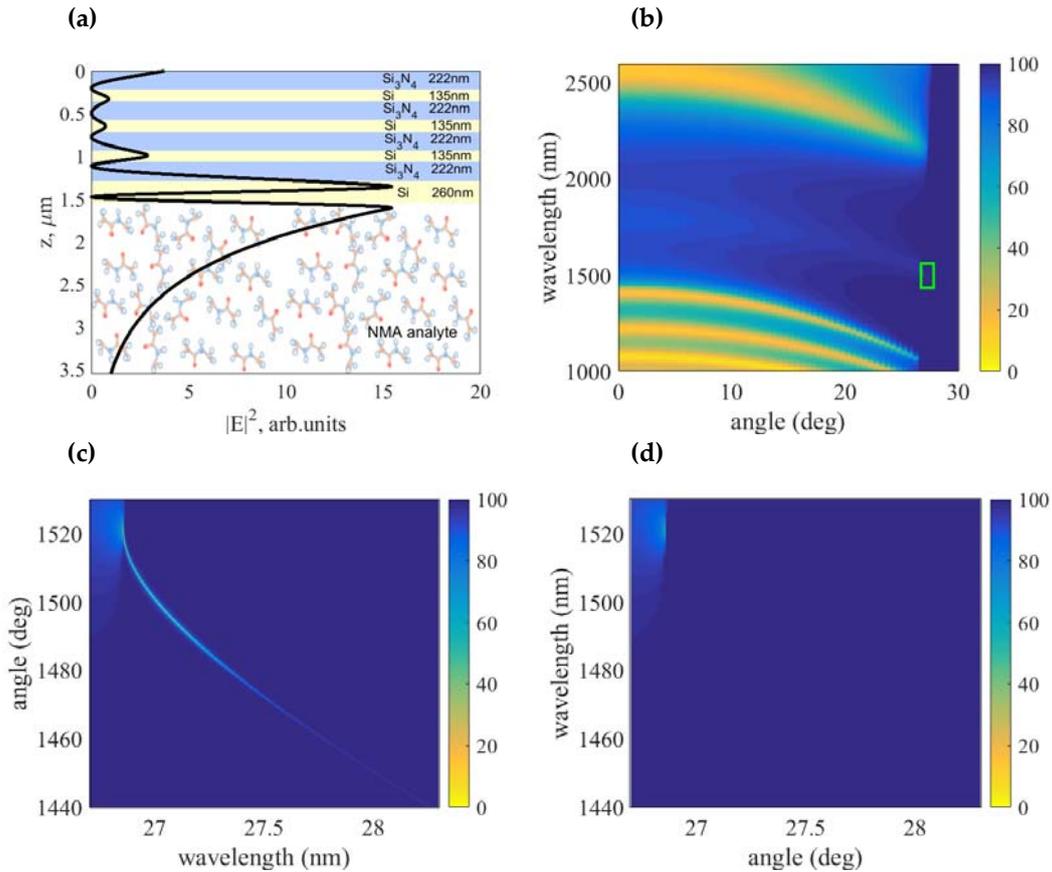

Fig. 2. (a) The PC structure scheme and the electromagnetic field distribution of the mode inside the PC structure and NMA analyte. The reflectance spectrum (incidence angle vs wavelength) of the PC-based structure for super-high Q sensing: (b) large scale, the position of the excited mode with respect to the PC bandgap and total internal reflection angle is shown by green rectangle, (c) magnified scale showing the ultrahigh-Q of the excited mode, (d) magnified scale showing the disappearance of the resonance in the case of zero absorption.

Figure 2b shows the reflectance at the incident angle vs. wavelength spectrum in non-magnetic sensor configuration. Ultranarrow resonance (also shown in Fig. 2c in the magnified scale) is in the bandgap of the PC structure and also below the total internal reflection line ($\theta$=26.65 deg). Therefore, the designed structure provides the waveguide propagation of light inside the structure due to the total internal reflection from the analyte on the one side, and a photonic BG on the other side. The structure is designed to provide the near-cutoff conditions, so that the light intensity distribution inside this bottom waveguide Si layer and in the neighboring regions is highly asymmetric and efficiently penetrates inside the analyte, see Fig.2a.

The bright line in Fig. 2c depicts the position of the resonance associated with the weak absorption peaks of NMA molecules. One can see that the proposed design guarantees the existence of the high-Q resonance in the spectral range from 1.48 um to 1.52 um, which coincides with the NMA absorption line of the N-H vibrational mode. In the shorter wavelengths the addressed resonance becomes shallow and gradually disappears. Therefore, even weak absorption peak at 1.495 um with extinction coefficient as small as $n''$=10$^{-4}$ (typical to the NMA molecules) causes a noticeable dip in the reflection spectra of light at an angle of incidence of 27.1°.

In order to reveal the impact of small extinction resonance of the studied molecules on the reflectance resonance, we have numerically compared optical response of the sensing structure in two cases, with ($n''=10^{-4}$, Fig.2c) and without ($n''=0$, Fig. 2d) absorption of the analyte. If we set the extinction coefficient of the analyte to zero and leave all other parameters and dielectric properties **including refraction indices** of the materials unchanged, the resonance in the reflection spectra clearly seen in Fig.2c disappears in Fig.2d. The absence of the resonance in Fig. 2d shows that the resonant dip in the reflection emerges due to the absorption properties of the NMA molecules. This behavior is explained by the coupled back radiation into the reflected light, which in lossless structure is 100% (also see Appendix C). This is a key fact for the addressed sensor of the NMA molecules.

## 4. ULTRA-NARROW MAGNETO-OPTICAL RESONANCE IN THE PC-BASED MAGNETIC NANOSTRUCTURE

An important feature of the magneto-optic (MO) effects is the fact that they are strongly dependent on the dielectric properties of materials, including the absorption. Moreover, it has been reported earlier that even minute changes of the permittivity tensor could be sufficient for the perceptible changes in the MO effects spectra in specially designed structures providing the novel approach to optical sensing [18, 29, 30].

Usually, the transverse magneto-optical Kerr effect (TMOKE) measured as the relative change of the reflection coefficient under structure remagnetization is addressed in MO sensing structures.

TMOKE could be observed only in *p*-polarized light, thus in contrast to the non-magnetic PC nanostructure, here we tune the PC parameters to excite TM-type mode. PC consisting of 20 pairs of $Ta_2O_5$ (255 nm) / Si (373 nm) layers is terminated with magnetic cerium-substituted yttrium iron-garnet (CeYIG), deposited to provide MO response. The PC nanostructure under consideration is placed inside the reversible magnetic field directed along the plane of the PC layers and orthogonally to the wavevector of the incident light. The scheme of the addressed nanostructure and the applied magnetic field H is given in Figs. 1 and S4. One of the equilateral $TiO_2$ prism side facets is illuminated by *p*-polarized light. The reflected light is detected on the other side facet of the prism.

The proposed design of PC nanostructure demonstrates narrow optical resonance at the wavelength 1.495 um for the incident angle of 40.43° [Fig. 3a]. This resonance is related to the excitation of TM-type quasi-surface mode. Applied external magnetic field affects that mode and thereby changes the optical response of the structure. As a result, the resonant enhancement of TMOKE is observed [Fig. 3b].

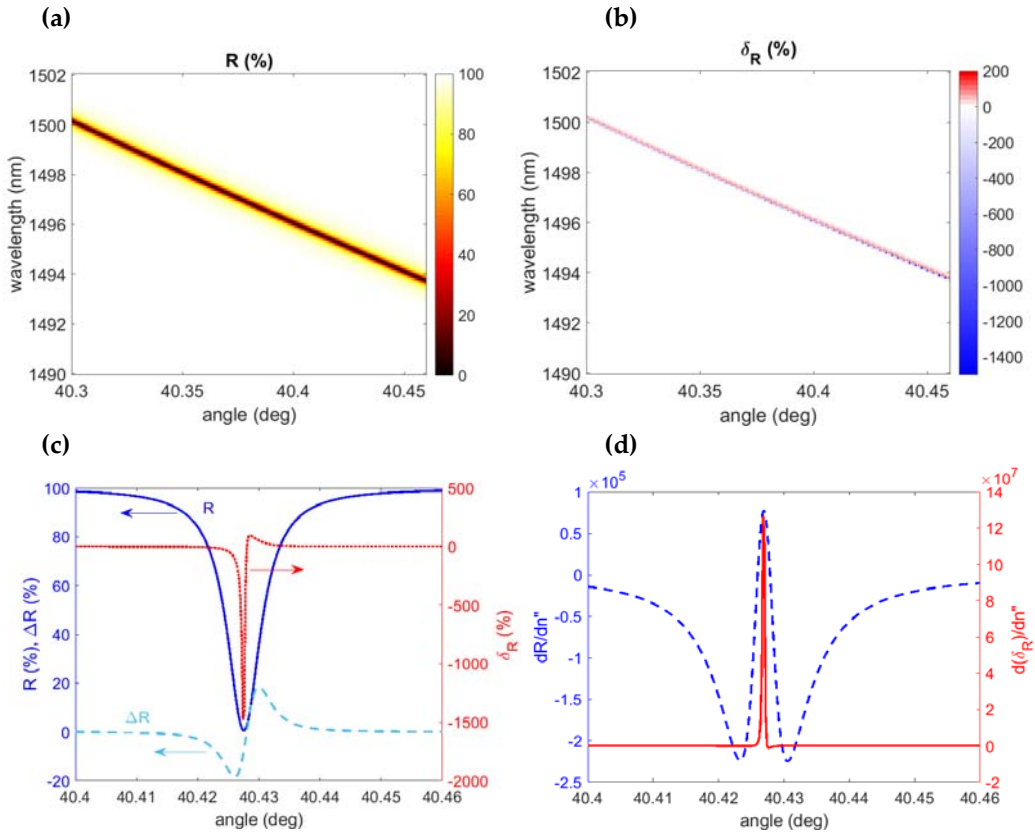

Fig. 3. Magnetic sensing structure with TM-polarized quasi-surface mode: (a) reflectance (*R*) and (b) TMOKE magnified scale wavelength vs angle plot showing the ultrahigh-Q of the excited mode and corresponding enhancement of the MO response; (c) *R* (solid line), *ΔR* (dashed line) and TMOKE $δ_R$ (dotted line) angular spectra at the resonance wavelength 1.495 um, (d) angular spectra of sensitivity of optical and MO response to the extinction coefficient of the analyte, $dR/dn''$ and $d(dR/R)/dn''$.

Let us analyze in detail how the reflectance spectra, *ΔR*, and TMOKE, $δ_R$, behave in the presence of a weakly absorbing medium, the NMA molecules for the case. The angular spectra of the reflection *R*, *ΔR* and a ratio $δ_R$ are given in Fig. 3c. One can see that the designed structure has a clear dip in the reflection spectra. The width of the resonance in *ΔR* spectrum is comparable with the resonance in the reflection spectra. However, the resonance in TMOKE spectrum, $δ_R$, is significantly narrower and has a higher Q-factor. This feature of the TMOKE resonance was earlier reported to enhance the sensitivity of the sensors measuring the changes of the real part of the medium refractive index.

We argue that high Q-factor of TMOKE resonance together with the enhanced magnitude up to 1000% is also responsible for the amplification of the sensitivity to the imaginary part of the medium refractive index. Therefore, the small absorption peaks of the material in NIR could be detected and the presence of NMA molecules in analyte could be measured with greater precision.

Figure 3 shows the sensitivity of the reflectance and TMOKE spectra to the variation of the extinction coefficient. The dramatic increase of the magneto-optical sensitivity more than one order of magnitude compared to the sensitivity of the optical sensor is clearly observed near the resonance of the quasi-surface electromagnetic mode.

## 5. DISCUSSION

To conclude, we have theoretically explored for the first time the magnetooptical effects on the molecular vibrational overtones in near-IR. We proposed two types of the PC sensor configurations that provide an ultra-narrow super-high-Q resonances associated with the higher harmonics of the vibrational transitions of organic molecules NMA. In contrast to fundamental vibrations that have rather good signal-to noise ratio, but lay in the mid-IR, higher harmonics emerge in near-IR region, but have very low signal-to-noise ratio. Therefore, the overtones are challenging to detect.

In the proposed PC nanostructures we take an advantage of the super-high-Q modes in PC-based structures and MO effects and show detectability of NMA molecules. By exciting the quasi-surface modes, well-known for their extreme sensitivity to the materials properties, the weak absorption peaks corresponding to the higher harmonics of the N-H vibrational transitions can be effectively detected. We study both magnetic and non-magnetic PC nanostructure sensors. The main advantage of the non-magnetic PC nanostructure sensor is the small number of the layers in the PC structure. The design presented in this paper contain just 4 pairs and one additional Si layer. In the presence of NMA molecules in analyte the reflection spectrum of the proposed non-magnetic PC nanostructure demonstrates an unambiguous change of profile that can be easily detected in experiment. Moreover, the magnetic PC nanostructure opens the possibility to detect the presence of NMA molecules in analyte with even greater precision.

Magnetic sensor demonstrates several orders of magnitude higher sensitivity as compared to non-magnetic sensing due to the measurements of super-high-Q magnetooptical resonances. It is important, that not only the Q-factor of the resonance increases, but also the magnitude of the measured value (i.e. TMOKE instead of $R$), which makes the proposed structure very promising for vibrational overtone sensing applications. It should be noted that this sensor requires very accurate fabrication. Specifically, the fabrication tolerances of the layer thicknesses should not exceed 2 nm to provide such sharp and deep resonances. The tolerance of the structure to the deviations of the parameters is described in Appendix A. This fabrication restriction is possible and accessible with present technologies of thin-film deposition. It is important to note that the, magnetooptical sensing configuration has the advantages of significantly improved signal-to-noise ratio (SNR). Also, since the measurement of the relative variation of the reflectance does not require precise normalization of the intensity, it also helps to avoid any spurious interference present in the experimental scheme. This is very important for the development of highly sensitive spectroscopic tools based on high-Q magneto-photonic sensors of high SNR.

Therefore, the presented sensing structure allows for the precise detection of small absorption peaks of the material in NIR, which are the 'fingerprints' corresponding to the certain molecules and bounds.

## 6. MATERIALS AND METHODS

Photonic crystal is designed to provide an ultranarrow resonance of quasi-surface electromagnetic wave in accordance with the impedance approach described, for instance, in Refs. [35, 36]. The parameters of the photonic crystal are chosen to provide the optical impedances same for both sides of the PC/analyte interface. This method is based on Fresnel formulas and allows for the direct numerical calculation of optical properties of the structure, including far-field (e.g. reflectance) and near-field (electromagnetic field distribution) response.

We design the non-magnetic photonic crystal nanostructure [see Fig. 2a] which is composed of 4 pairs of [silicon]/[silicon nitride] layers. The thickness of Si layers in PC is 135 nm, and $Si_3N_4$ layers are 222 nm-thick each. Besides that, there is a bottom Si layer with the thickness of $h$=260 nm between PC and the analyte. To introduce the s-polarized light into the PC structure at the required incidence angle one can use an equilateral silicon (Si) prism with the base angles of about 27.1° depending on the angular position of the resonance in the reflection spectra.

In addition, we propose the magnetic PC nanostructure composed of 20 pairs of [$Ta_2O_5$]/[Si] layers and additional magnetic layer of cerium-substituted yttrium iron garnet (CeYIG). The order and thickness of the layers have been optimized to achieve minimal values of both reflection coefficient $R$ and impedance $Z$ of the structure. The best results were obtained for the PC nanostructure given in Fig. S4. 255 nm-thick $Ta_2O_5$ layers alternate with 373 nm-thick Si layers. The CeYIG film should be 50 nm thick.

The fabrication technologies required for making the proposed PC nanostructures are well-developed and widely applied. For instance, the similar PC structures have been experimentally studied in Refs. [28-33, 36-38] to mention just a few. Note, that the typical fabrication precision is about 0.5nm [36-38], so the fabrication tolerance of 2nm required for the proposed PC nanostructures can be easily provided.

The refractive indices of the materials at the wavelength 1.495 um were taken as the following $n_{Si}$=3.4804 at T=293K [39], $n_{Si3N4}$=1.9978 [40], $n_{Ta2O5}$=2.0591 [41], $n_{TiO2}$=2.4548 [42], $n_{NMA}$=1.5712 and $n''_{NMA}$=$10^{-4}$ [4, 7], $n_{CeYIG}$=4.95 and gyration $g_{CeYIG}$=$10^{-2}$ [see, for instance, 43].

**Funding Information.**
**The study of absorption detection in NMA was supported by** Israeli Innovation Authority-Kamin Program (62045 Year 2). The study of non-magnetic ultrahigh-Q structure was supported by Russian Foundation for Basic Research (RFBR) (19-02-00856_a). The study of magnetic ultrahigh-Q all-dielectric structure performed by D.O.I. was supported by RSF, project no. 18-72-00233.

**Acknowledgment**. V.I.B. and O.V.B. acknowledge support from the Foundation for the advancement of theoretical physics BASIS.

See Appendices for supporting content.

## APPENDIX A: TOLERANCE TO THE LAYER THICKNESS AND REFRACTIVE INDEX

The high-Q resonance in the proposed structures is very sensitive to the deviations of the widths and refractive indices of the layers. In order to study the structure tolerance in details, we have performed the calculations where the thicknesses $w_j$ and the refractive indices $n_j$ of the structure layers were disturbed by randomly distributed values of the certain magnitude: $w_j+\delta w_j$, and $n_j+\delta n_j$. Fig. S1 shows the relative magnitude $I/I_0$ of the quasi-surface wave intensity in the disturbed structure normalized on the surface wave intensity in the non-disturbed structure $I_0$, with respect to the magnitude of $\delta w_j$ and $\delta n_j$ averaged over 100 randomly generated deviations.

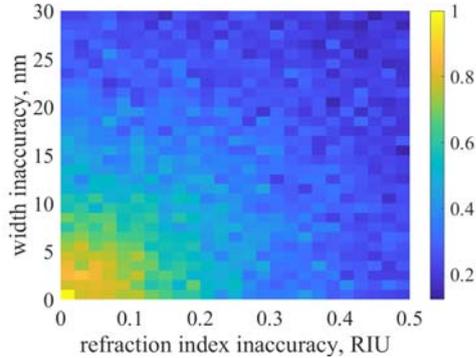

Fig. S1. Relative magnitude $I/I_0$ of the quasi-surface wave intensity in the disturbed structure normalized on the surface wave intensity in the non-disturbed structure $I_0$, with respect to the magnitude of $\delta w_j$ in nm and $\delta n_j$ in refractive index units (RIU) deviations from the designed valyes averaged over 100 randomly generated deviations.

According to Fig. S1, the accuracy of the layer deposition provides $\delta w_j$<10 nm and $\delta n_j$<0.15 RIU provides less than 2 times decrease of the quasi-surface wave magnitude compared to the ideal case. Less than 2 nm deviation of the layer widths and less than 0.05 RIU deviation of the refractive indices guarantees the difference of about 10% compared to the ideal structure.

In the Fig. S2 we plot the evolution of the resonances as a function of the bottom silicon layer height $h$, which is a guiding layer for quasi-surface wave. The narrowest and deepest resonance is provided by the thinnest silicon layer, $h=255$ nm. This occurs at about 26.8° (solid black curve). Even though the thicker layers of $h=260$ nm and $h=265$ nm don't provide such narrow resonances, the corresponding dips in the reflection spectra are still sufficient for the effective detection of NMA molecules. Moreover, with the growth of $h$ the angular position of the resonance moves toward larger values of the incident angle. Naturally, Si layer thickness of *255 nm* is preferable, but in many cases the standard accuracy of the fabrication admits the thickness tolerances of about a few nanometers. Thus, it is crucial to guarantee that a proposed design of the PC nanostructure still has high-Q resonances even in case of small deviation from dictated by the design thickness.

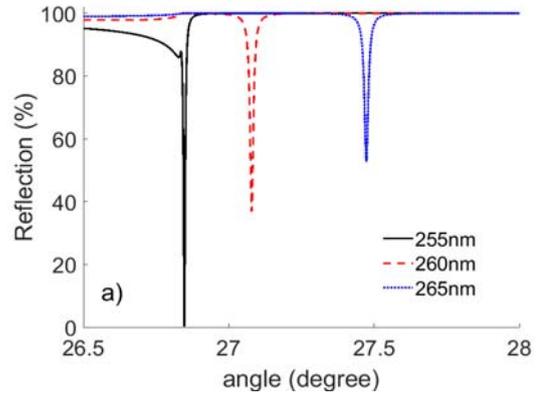

Fig. S2. Normalized reflectance (in % of light energy reflected by the structure normalized to the incident energy) vs angle of incidence (degrees) of super-high Q resonance at the wavelength of the incident light is 1.495 um. Resonant curves corresponding to different thicknesses of the bottom Si layer: 255 nm (black solid curve), 260 nm (red dashed curve), and 265 nm (blue dotted curve).

## APPENDIX B: ANGULAR SPECTRA FOR PC-BASED STRUCTURE WITH QUASI-GUIDED MODES

There are two approaches how to obtain angular spectra in sensing measurements. The first one is based on utilization of the collimated light with the angular width much smaller than the dip width, 0.01 deg namely in the case of PC-based ultrahigh-Q structures. The angle of incidence is varied by the mechanical motion of the sample with prism, and the corresponding motion of the light detector. This scheme would provide the reflectance curves shown in Fig. S3 for the cases of absorbing and 'non-absorbing' NMA material.

The second approach is utilization of the light beam with higher angular width, focused at the sample surface via the lens. After reflection from the sample, the light is collimated via the prism and detected by CMOS matrix so that each pixel corresponds to the certain angle of incidence. This approach allows for the simultaneous measurement of the angular spectra. However, it should be noted that the interference of the light, reflected directly from the sample, and the light, transferred to the waveguide mode and the coupled back, will take place. This will cause the typical interference pattern in angular spectrum of the reflected light (similar to the one observed in [27]), however, the excitation of long-range propagating mode will still result in the dip corresponding to its absorption.

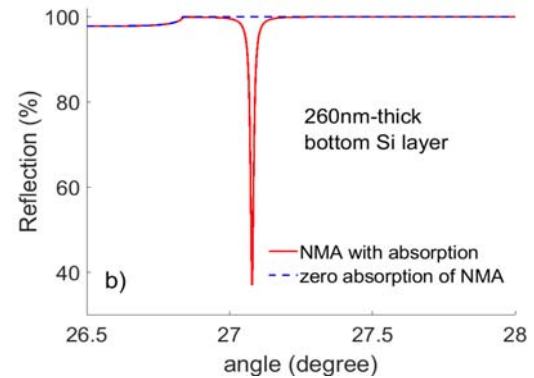

Fig. S3. Normalized reflectance (in % of light energy reflected by the structure normalized to the incident energy) vs angle of incidence (degrees) of super-high Q resonance at the wavelength of the incident light is 1.495 um. Resonant curves in the cases of absorbing ($n''=10^{-4}$, red solid curve) and non-absorbing ($n''=0$, blue dashed curve) analyte material with the same real part of the refractive index.

## APPENDIX C: SCHEME OF SENSING WITH MAGNETIC PC-BASED STRUCTURE WITH QUASI-GUIDED MODES

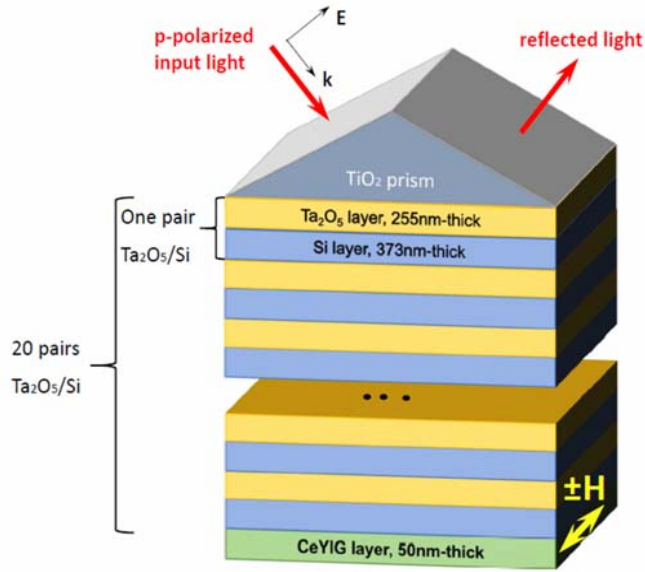

Fig. S4. The scheme of the magnetic PC nanostructure in the external magnetic field *H*. Input and reflected light at 1.495 um is shown with red arrows.

In the Fig. S4 on can see the scheme of the proposed magnetic PC nanostructure for weak absorption peak sensing addressed in Section 4. PC nanostructure consists of 20 pairs of $Ta_2O_5$ (255 nm)/Si (373 nm) layers and is terminated with magnetic cerium-substituted yttrium iron-garnet (CeYIG), deposited to provide MO response. The structure under consideration is placed inside the reversible magnetic field ***H*** directed along the plane of the PC layers and orthogonally to the wavevector of the incident light. Incident light is *p*-polarized to provide the excitation of TM-mode inside the nanostructure. The reflected light is detected on the other side facet of the prism.